\documentclass[preprintnumbers,amsmath,amssymb]{revtex4}
\usepackage{color}
\usepackage{graphicx}
\usepackage{epsfig}
\usepackage{epstopdf}
\usepackage{ragged2e}
\usepackage{mathrsfs}
\usepackage{csquotes}
\usepackage{paralist}
\usepackage{amsthm}
\usepackage[inline]{enumitem}
\usepackage{bm}
\newtheorem{corollary}{Corollary}[section]

\newtheorem{theorem}{Theorem}[section]
\newtheorem{proposition}{Proposition}[section]
\newtheorem{lemma}{Lemma}[section]

\begin{document}

\title{The closeness of the Ablowitz-Ladik lattice to the Discrete Nonlinear Schr\"odinger equation}
\author{Dirk Hennig}
%  \affiliation{Institut f\"ur Physik, Humboldt
% Universit\"at zu Berlin, Newtonstr. 15, 12489 Berlin, Germany}
\author{Nikos I. Karachalios}
\affiliation{Department of Mathematics,  University of Thessaly, 35100, Lamia, Greece}
\author{Jes\'{u}s Cuevas-Maraver}
%\email[Email: ]{jcuevas@us.es}
\affiliation{Grupo de F\'{i}sica No Lineal, Departamento de F\'{i}sica Aplicada I,
	Universidad de Sevilla. Escuela Polit\'{e}cnica Superior, C/ Virgen de \'{A}frica, 7, 41011-Sevilla, Spain \\
	Instituto de Matem\'{a}ticas de la Universidad de Sevilla (IMUS). Edificio Celestino Mutis. Avda. Reina Mercedes s/n, 41012-Sevilla, Spain}
%\author{Dirk Hennig}
%\author{Dirk Hennig}
%%	\affiliation{Department of Mathematics, University of the Aegean, Karlovassi, 83200
%%		Samos, Greece}
%\author{Nikos I. Karachalios}
%\author{Jes\'{u}s Cuevas-Maraver}
% \affiliation{Institut f\"ur Physik, Humboldt
% Universit\"at zu Berlin, Newtonstr. 15, 12489 Berlin, Germany}
\date{\today}
\begin{abstract}
While  the Ablowitz-Ladik lattice is integrable, the Discrete Nonlinear Schr\"odinger equation, which is more significant for physical applications, is not. We prove closeness of the solutions of both systems in the sense of a ``continuous dependence" on their initial data in the $l^2$ and $l^{\infty}$ metrics. The most striking  relevance of the analytical results is that small amplitude solutions of the Ablowitz-Ladik system persist in the Discrete Nonlinear Schr\"odinger one.  It is shown that the closeness results are also valid in higher dimensional lattices as well as for generalised nonlinearities.   For illustration of the applicability of the approach, a brief  numerical study is included,  showing that when the 1-soliton solution of the Ablowitz-Ladik system is initiated in the Discrete Nonlinear Schr\"odinger system with cubic and  saturable nonlinearity, it persists for long-times. Thereby excellent agreement of the numerical findings with the theoretical predicti
ions is obtained.
\end{abstract}
%
%\pacs{05.45.-a, 63.20.Pw, 45.05.+x, 63.20.Ry}
\maketitle

\noindent
\section{Introduction}
The Discrete Nonlinear Schr\"odinger equation (DNLS)
\begin{equation}
i\dot{\phi}_n+{\kappa\nu}\left(\phi_{n+1}+\phi_{n-1}\right)+\gamma
|\phi_n|^2\phi_n=0,\,\,\,n\in {\mathbb{Z}},\;\;\gamma>0,\label{eq:DNLS0}
\end{equation}
is one of the most important nonlinear lattice systems \cite{HennigTsironis},\cite{Kevrekidis},\cite{Eil},\cite{DNLS}. It appears as a fundamental, inherently discrete model in a great variety of physical contexts. The study of its dynamics has been an exciting topic of research as it deals with such diverse physical and biological phenomena  as wave motion in coupled nonlinear waveguides, dynamics of modulated waves in nonlinear electric lattices, localisation of electromagnetic waves in photonic crystals, energy localisation in discrete condensed matter and biological
systems  and  the dynamics of Bose-Einstein Condensates \cite{BEC1}, to mention a few. In Eq. \eqref{eq:DNLS0}, $\kappa>0$ is a discretisation parameter, while the sign of the parameters $\nu$ and $\gamma$ renders the DNLS as focusing (same sign) or defocusing (opposite sign).

A crucial difference from its continuous limit $\kappa\rightarrow\infty$ leading to  the cubic  Nonlinear Schr\"odinger equation (NLS)
\begin{eqnarray}
i\partial_t u + \nu u_{xx}+\gamma |u|^{2}u=0,\;\; x\in\mathbb{R},\label{eq:NLS0}
\end{eqnarray}
is that the DNLS \eqref{eq:DNLS0} is a non-integrable discretisation of the  integrable partial differential equation \eqref{eq:NLS0}. This is not the case for another  discretisation of NLS \eqref{eq:NLS0},
known as the
Ablowitz-Ladik equation (AL) \cite{AL}, \cite{AL2}, \cite{Herbst},
\begin{equation}
i\dot{\psi}_n+\kappa\nu(\psi_{n+1}+\psi_{n-1}) +\mu\,|\psi_{n}|^2(\psi_{n+1}+\psi_{n-1})=0,\,\,\,n\in {\mathbb{Z}},\;\;\mu> 0.
%\in\mathbb{R}.
\label{eq:AL0}
\end{equation}
Like in (\ref{eq:DNLS0}) the sign of the parameters $\nu$ and $\mu$  determines whether the model is of focusing or defocusing kind.  Most importantly, the AL lattice is an integrable discretisation of \eqref{eq:NLS0}, as it was shown by the discrete version of the Inverse Scattering Transform \cite{AL}, and therefore it has an infinite number of conserved quantities. We remark that the AL is one of the few known completely integrable (infinite) lattice systems  admitting soliton solutions \cite{Ablowitz},\cite{Faddeev}. On the infinite lattice with vanishing boundary conditions, general solutions can be obtained \cite{AL2}. For example, for $\kappa=\nu=1$ and $\mu>0$, the one-soliton solution reads as
\begin{equation}
\begin{split}
\psi^s_n&=\frac{\sinh\beta}{\sqrt{\mu}}\mathrm{sech}\left[\beta(n-v_st)\right]\exp(-i(\omega t-\alpha n)),\\
\;\;\omega&=-2\cos \alpha \cosh \beta,\\
v_s&=2\beta^{-1}\sin \alpha \sinh \beta,\label{eq:one-soliton}
\end{split}
\end{equation}
with $\alpha \in [-\pi,\pi]$ and $\beta \in [0,\infty)$.
System (\ref{eq:AL0}) possesses the following conserved quantity
\begin{equation}
P_{\mu}=\sum_{n}\ln(1+\mu |\psi_n|^2)\label{eq:Pmu}.
\end{equation}

In \cite{Ver1}, the Inverse Scattering Transform method has been developed for nonvanishing boundary conditions and N-dark soliton solutions of the AL equation have been given in terms of the Casorati determinant in \cite{Maruno}. Yet in similarity with the integrable NLS \eqref{eq:NLS0}, another important class of solutions of the AL is the one of rational solutions which are discrete versions of the Peregrine soliton and the Kuznetsov-Ma breather \cite{akhm_AL}, \cite{akhm_AL2}.
%%
%: Applying when $\kappa=\nu=1$ and $\mu>0$ in \eqref{eq:AL0}, the transformation $\psi_n(t)\rightarrow \psi_n(t)\exp[2i(q^2-1)t]$, brings it into the form
%\begin{equation}
%i\dot{\psi}_{n}+\left(\psi_{n+1}-2\psi_{n}+\psi_{n-1}\right)%
%+\mu|\psi_{n}|^{2}\left(\psi_{n+1}+\psi_{n-1}\right)-2q^{2}\psi_{n}=0,
%\label{eq: AL01}
%\end{equation}
%where the parameter $q$ fixes a  background amplitude. The AL \eqref{eq: AL01} possesses a discrete counterpart of the Kuznetsov-Ma (KM) breather of the NLS \eqref{NLS0}, a time-periodic
%solution given explicitly by
%%
%\begin{equation}
%\psi_{n}(t)=q\frac{\cos{\left(\omega t+i\theta\right)+G\cosh{\left(r n\right)}}}%
%{\cos{\left(\omega t\right)+G\cosh{\left(r n\right)}}},
%\label{km_exact}
%\end{equation}
%with frequency $\omega$ (related to the period of the solution via $T=2\pi/\omega$),
%$\theta=-\arcsinh{\left(\omega\right)}$, $r=\arccosh{\left(\left[2+\cosh{(\theta)}\right]/3\right)}$
%and $G=-\omega/\left(\sqrt{3}\sinh{\left(r\right)}\right)$, \cite{akhm_AL}, \cite{akhm_AL2}.
%%
%The long period limit $\omega\rightarrow 0$ of \eqref{km_exact} results in the discrete counterpart of the Peregrine rogue wave of the NLS \eqref{NLS0},
%\begin{equation}
%\psi_{n}(t)=q\left[1-\frac{4(1+q^2)(1+4iq^2t)}{1+4n^2q^2+16q^4t^2(1+q^2)}\right].
%\label{prw_exact}
%\end{equation}

Strictly speaking, all the aforementioned analytical solutions  exist only for the AL. In the case of the soliton solution \eqref{eq:one-soliton} of the AL this is manifested in the fact that it exhibits continuous translation symmetry and  possesses  a band of velocities for each $\beta$, which allow it to travel along the lattice. This is not the case for the DNLS for which a localised state can be pinned due to the Peierls-Nabarro barrier \cite{PN1},\cite{PN2}, \cite{PN3}.

 Therefore, the question of the persistence of solitary wave dynamics and the existence of localised structures in non-integrable  lattices, such as the DNLS \eqref{eq:DNLS0}, has attracted tremendous interest. As a milestone in this context, we recall the construction of localised in space, time periodic or time-quasiperiodic solutions of lattice dynamical systems, including  the DNLS as special case \cite{MackayAubry}, \cite{Aubry},\cite{MJ}, starting from the anti-continuous limit $\kappa\rightarrow 0$.  For key works regarding numerical computations of discrete solitons we refer to \cite{ChrisFl},\cite{AubyMarin}. Other seminal results on the existence of nonlinear localised modes of the DNLS equations have been gained by nonlinear analysis methods, in particular variational ones, establishing the existence of localised structures as critical points of suitable functionals \cite{Wein99},\cite{Pan1},\cite{Pan2},\cite{Pan3},\cite{Pan4}.  Important extensions concern the
 existence of more complex structures in higher dimensional set-ups, such as the discrete vortex solutions, see \cite{DNLS} and references therein.
The crucial issue of existence and stability of travelling solitons in DNLS lattices has been investigated by a combination of analytical and numerical methods verifying in many cases the robustness of the discrete localised structures under perturbations, \cite{Peli0},\cite{Jesus1},\cite{Peli1},\cite{Peli2}. In this context we also refer to the reviews \cite{FlachWillis}, \cite{reviewsC}, \cite{PanosImaRev}.

In the present work, we investigate  a persistence/existence problem, by  examining in the sense of  ``continuous dependence'', the closeness of the solutions of the DNLS and the AL for close enough initial data. That is, the following question is investigated here: {\em assuming that the initial data of the DNLS \eqref{eq:DNLS0} and the AL \eqref{eq:AL0} are sufficiently close in a suitable metric, do the associated solutions remain close for sufficiently long times?}
We argue that answering this question is important because of the following reasons:
\begin{itemize}
	\item
 Whereas, it is natural to expect, at least in some cases of parametric regimes, that sufficiently weak non-integrable perturbations (e.g. stemming from gain/loss or forcing terms or higher order terms) lead to solutions staying  close to those of the underlying integrable (core) system dynamics, in the case of the DNLS and AL lattices, there is not such a limiting connection between the  systems.
 \item  An affirmative answer to the question above will establish that the already diverse  dynamical features  of the DNLS itself  are  even further enriched as then the DNLS closely share such solutions that are provided by the functional form of the analytical solutions of the AL-lattice (at least) for small amplitudes. From this perspective, not only the soliton solutions but also the discrete rational solutions are relevant.
\end{itemize}
In our aim to answer the above question,  we proceed by analytically proving that \textit{at least under certain smallness conditions on the initial data of the DNLS and the AL lattices, the corresponding solutions remain close for all times.} To be precise, we state the result for the infinite lattice with vanishing boundary conditions
\begin{equation}
\label{vanbc}
\lim_{|n|\rightarrow\infty}\phi_n=\lim_{|n|\rightarrow\infty}\psi_n=0.
\end{equation}
Hence, the natural phase space for the systems is the Hilbert-space of the square-summable sequences
\begin{equation}
\label{sl2}
l^2=\left\{ \phi=(\phi_n)_{n \in {\mathbb{Z}}}\,\in {\mathbb{C}}\,\,\,\vert \, || \phi||_{l^2}=\left(\sum_{n }|\phi_n|^2\right)^{1/2}\right\}.
\end{equation}
Consider then, the initial conditions for the DNLS \eqref{eq:DNLS0} and AL \eqref{eq:AL0}
\begin{equation}
\label{inDNLS}
\phi_n(0)=\phi_{n,0},\,\,\,n\in {\mathbb{Z}},
\end{equation}
and
\begin{equation}
\label{inAL}
\psi_n(0)=\psi_{n,0},\,\,\,n\in {\mathbb{Z}},
\end{equation}
respectively. The main result of the paper is the following
\begin{theorem}
\label{Theorem:closeness}
Consider the DNLS equation \eqref{eq:DNLS0}. We assume that  for every  $0<\epsilon<1$,
the initial conditions \eqref{inDNLS} of the DNLS \eqref{eq:DNLS0}  and the initial conditions \eqref{inAL} of the AL \eqref{eq:AL0} satisfy:
	\begin{eqnarray}
	\label{eq:distance0}
	|| \phi(0)-\psi(0)||_{l^2}&\le& C_0\, \varepsilon^3,\\
	\label{eq:distance01}
	|| \phi(0)||_{l^2}&\le& C_{\gamma,0}\, \varepsilon,\\
	\label{eq:Pmu0}
	P_{\mu}(0)&=&\sum_{n}\ln(1+\mu|\psi_n(0)|^2)\le C_{\mu,0}\,\varepsilon^2
%	\ln(1+(C_{\mu,0}\,\varepsilon)^2),
	\end{eqnarray}
for some constants $C_0, C_{\gamma,0},\,C_{\mu,0}>0$.	
Then, for arbitrary finite $0<T_{\small{f}}<\infty$, there exists a constant $C=C(\gamma,\mu, C_0, C_{\mu,0},C_{\gamma,0},T_{\small{f}})$, such that the corresponding solutions
for every $t\in [0,T_{\small{f}}]$, satisfy the estimate 	
	\begin{equation}
||y(t)||_{l^2}=|| \phi(t)-\psi(t)||_{l^2}\le C \varepsilon^3.\label{eq:boundy}
	\end{equation}
\end{theorem}
For the proof of Theorem \ref{Theorem:closeness} we use suitable estimates for the  solutions of the AL lattice based on its \textit{deformed power or norm}, and energy arguments for the difference of  solutions of the systems. It also makes essential use of the global existence of solutions for both lattices ensured by considering a physically relevant variant of a DNLS system which combines both systems studied first in \cite{Cai94} (the so-called  Salerno model, see \cite{Salerno}, \cite{Cai94}).

An immediate consequence follows from Theorem \ref{Theorem:closeness}, due to embedding $||y||_{l^{\infty}}\leq ||y||_{l^{2}}$ which holds for every $y\in l^2$ and can be applied to the estimate \eqref{eq:boundy}.

\begin{corollary}
\label{Corollary:closeness2}
 %When $\phi(0)=\psi(0)$,
 Under the assumptions \eqref{eq:distance01}-\eqref{eq:Pmu0}, for every $0<\epsilon<1$ and $t\in [0,T_{\small{f}}]$, the maximal distance $|| y(t)||_{l^\infty}=\sup_{n\in {\mathbb{Z}}}|y_n(t)|=\sup_{n\in {\mathbb{Z}}}|\psi_n(t)-\phi_n(t)|$ between individual units of the systems satisfies the estimate
\begin{equation}
||y(t)||_{l^\infty}\le \tilde{C} \varepsilon^3,\label{eq:boundy1}
\end{equation}
for some constant $\tilde{C}=\tilde{C}(\gamma,\mu, C_{\mu,0},C_{\gamma,0},T_{\small{f}})$.
\end{corollary}
% Theorem \ref{Theorem:closeness2} is proved with an alternative method which  makes use of the Fourier transform of the local distance function of the solutions.

Theorem \ref{Theorem:closeness} establishes the closeness of the solutions to the AL and the DNLS as $\varepsilon \rightarrow 0$, with an explicit expression for the associated constant $C$ in dependence on the parameters of both systems.
%interestingly, we identify that the quantifications of the constants $C$ and $\tilde{C}$ derived from both approaches of the proofs are consistent (see remark \ref{comcon}).
Its main application is that it rigorously justifies that at least small amplitude  localised structures provided by the analytical solutions of the integrable AL-lattice persist in the DNLS lattice. In other words, the DNLS lattice admits small amplitude solutions of the order ${\cal{O}}(\varepsilon)$, that stay ${\cal{O}}(\varepsilon^3)$-close to the analytical solutions of the AL for any  $0<T_{\small{f}}<\infty$. In this regard, the analytical arguments show that the growth of the distance $||y(t)||_{l^2}$ is uniformly bounded for any $\epsilon>0$ and $t\in (0,\infty)$ as
\begin{eqnarray}
\label{gr}
\frac{d}{dt}||y(t)||_{l^2}\leq M\,\varepsilon^3,
\end{eqnarray}
($M$ depends on the parameters and initial data but not on $t$), and consequently, the distance of solutions growths at most linearly
for any $t\in (0,\infty)$, as
\begin{eqnarray}
\label{gr1}
||y(t)||_{l^2}\leq M\, t\,\varepsilon^3.
\end{eqnarray}
%In the case of Corollary \ref{Theorem:closeness2},
%%where $\phi(0)=\psi(0)$,
%we have
%\begin{eqnarray}
%\label{gr2}
%||y(t)||_{l^{\infty}}\leq \tilde{M}\,t\,\varepsilon^3,\;\;\forall\;\;t\in (0,\infty).
%\end{eqnarray}
In a similar context, we refer also to the linear time-growth estimates for the relevant distant function between the solutions of the complex Ginzburg-Landau pde and the NLS pde, when the inviscid limit of the former is considered \cite{JWU}, which can even grow exponentially \cite{OG}.

The closeness result of Theorem \ref{Theorem:closeness}  can be extended to other important cases of DNLS systems. These include the DNLS with generic power-nonlinearity $F(z)=|z|^{2\sigma}z$ for $\sigma>0$ and saturable nonlinearities of the form $F(z)=\frac{z|z|^2}{1+|z|^2}$ and $F(z)=\frac{z}{1+|z|^2}$. The extensions may consider higher-dimensional lattices $\mathbb{Z}^N$, $N\geq 1$.  Note that for  generalisations of the AL lattice in $\mathbb{Z}^2$, analytical localised solutions have been constructed \cite{Kim2},\cite{2DAL4},\cite{2DAL2},\cite{2DAL3}.

To corroborate our analytical results we include the results of a  numerical study treating the example of the soliton solution \eqref{eq:one-soliton} when launched on DNLS lattice with the cubic nonlinearity \eqref{eq:DNLS0}, and the DNLS with saturable nonlinearity. The numerical findings are in excellent agreement with the theoretical results.

%A first application concerns the discrete soliton \eqref{eq:one-soliton}, see Corollary \ref{corsol}.

The presentation of the paper is as follows: Section  \ref{secII} recalls some basic properties of the DNLS and the AL lattices, focusing on their conserved quantities and  auxiliary results that will aid the main proofs. In Section \ref{secIII} we prove the global existence result for the extended Salerno model
%IN-DNLS
of \cite{Salerno},\cite{Cai94}. Section \ref{SecIV} contains the proof of the main result Theorem \ref{Theorem:closeness}, while section \ref{Nlat} provides its extensions to higher-dimensional lattices and the saturable nonlinearities. In section \ref{SecNum} we present the results of the numerical study. Section \ref{SecCon} summarises the findings and provides a brief  plan for further relevant studies.
%\vspace*{0.5cm}

\section{Preliminaries}
\setcounter{equation}{0}
\label{secII}
For convenience, we set $\kappa=\nu=1$, without affecting the generality of the proofs, which are valid in either the focusing or the defocusing case. The AL \eqref{eq:AL0} can be derived from the Hamiltonian given by
\begin{equation}
 H=\sum_{n}
 %\left(
 \overline{\psi}_n(\psi_{n+1}+\psi_{n-1})
 %-\frac{1}{\mu}\ln(1+\mu|\psi_n|^2)\right),
\end{equation}
with the following deformed Poisson bracket
\begin{equation}
 \left\{\psi_m,\overline{\psi}_n\right\}=(1+\mu|\psi_m|^2)\delta_{m,n},\,\,\,\, \left\{\psi_m,{\psi}_n\right\}=\left\{\overline{\psi}_m,\overline{\psi}_n\right\}=0,
\end{equation}
yielding the equation of motion as
\begin{equation}
  \dot{\psi}_n=\left\{H,\psi\right\}.
\end{equation}
% The following conserved (deformed) norm of the AL determined by
% \begin{equation}
%  P_{\mu}=\sum_{n}\ln(1+\mu|\psi_n|^2),
% \end{equation}
% is of particular importance  for later studies.

The DNLS can be derived from the Hamiltonian
\begin{equation}
 H=\sum_{n}\left(\overline{\phi}_n(\phi_{n+1}+\phi_{n-1})-\frac{\gamma}{2}|\phi|^4\right),
\end{equation}
using the standard Poisson bracket and the equation of motion
\begin{equation}
i\dot{\phi}_n=\left\{H,\phi\right\}.
\end{equation}
For the DNLS the norm
\begin{equation}
\label{eq:Pgamma}
  P_{\gamma}=\sum_{n}|\phi_n|^2,
\end{equation}
is conserved.

The AL equation is completely integrable \cite{AL}, whereas its DNLS counterpart (\ref{eq:DNLS0}) is known to be nonintegrable \cite{DNLS},\cite{Herbst}.
Notice that in (\ref{eq:AL0}) and (\ref{eq:DNLS0}) the nonlinear terms are both of cubic order. However, they are \textit{markedly different} in the sense that, the nonlinear terms in (\ref{eq:AL0}) are of nonlocal nature compared to the local terms in (\ref{eq:DNLS0}).

In the case of the vanishing boundary conditions \eqref{vanbc}, the  functional space setting is based on the spaces of complex summable sequences
\begin{equation}
 l^p=\left\{ \phi=(\phi_n)_{n \in {\mathbb{Z}}}\,\in {\mathbb{C}}\,\,\, ||\phi||_{l^p}=\left(\sum_{n }|\phi_n|^p\right)^{1/p}\right\}.
\end{equation}
For any $\phi=(\phi_n)_{n \in {\mathbb{Z}}},\psi=(\psi_n)_{n \in {\mathbb{Z}}}\in l^2$ we consider the inner product
\begin{equation}
 (\phi,\psi)_{l^2}=\sum_{n \in {\mathbb{Z}}}\phi_n \overline{\psi}_n,
\end{equation}
where $\overline{\psi}$ denotes the conjugate of
$\psi_n$. With the associated
norm
\begin{equation}
 || \phi||_{l^2}^2=(\phi,\phi),
\end{equation}
$(l^2,(\cdot,\cdot),||\cdot||)$ is a complex Hilbert space.
We will use the continuous embeddings
\begin{eqnarray}
\label{embe}
 l^r\subset l^s,\,\,\,|| \phi||_{l^s}\le || \phi||_{l^r},\,\,\,1 \le r\le s \le \infty.\label{eq:embeddings}
\end{eqnarray}

The following auxiliary result will aid the ensuing studies.
\begin{lemma}
\label{lemaux1}
Let $\mu >0$. Assume that the initial condition \eqref{inAL} of the AL-lattice \eqref{eq:AL0} is such that
\begin{eqnarray}
\label{aux0}
P_{\mu}(0)=\sum_{n}\ln(1+\mu |\psi_n(0)|^2)< \infty.
\end{eqnarray}
Then, the corresponding solution of the AL lattice satisfies the estimate
% $\sum_{n\in{\mathbb{Z}}}|\psi_n(t)|^2=|| \psi(t)||_{l^2}<\infty ,\,\,\,\forall t\ge 0$. Moreover,
\begin{equation}
\mu || \psi(t) ||_{l^2}^2= \sum_{n}|\psi_n(t)|^2\le \exp(P_{\mu}(0))-1,\qquad \forall t\ge 0.\label{eq:psiconserved}
\end{equation}
\end{lemma}
\noindent{\bf Proof:} Using that the function
\begin{equation}
 f:\,{\mathbb{R}}_{+}\rightarrow {\mathbb{R}}_{+},\,\,\,x \mapsto \ln(1+\mu x),
\end{equation}
is continuous and bijective,
we write
\begin{equation}
\label{auxeq1}
P_{\mu}= \sum_{n}\ln(1+\mu|\psi_n|^2)=\sum_{n}|\lambda_n|^2.
\end{equation}
>From \eqref{auxeq1}, and by using the embedding \eqref{embe} for $s=2k$ and $r=2$, we get the estimate:
\begin{eqnarray}
 \mu \sum_{n}|\psi_n|^2&=&\sum_{n}\left(\exp(|\lambda_n|^2)-1\right)=\sum_{n}\,\left(\sum_{k=0}^{\infty}\frac{|\lambda_n|^{2k}}{k!}-1\right)\nonumber\\
 &=&\sum_{n}\,\sum_{k=1}^{\infty}\frac{|\lambda_n|^{2k}}{k!}=\sum_{k=1}^{\infty}\frac{1}{k!}\,\sum_{n}\left(|\lambda_n|^{2}\right)^k\nonumber\\
 &\le&\sum_{k=1}^{\infty}\frac{1}{k!}\,\left(\sum_{n}|\lambda_n|^{2}\right)^k=\sum_{k=1}^{\infty}\frac{P_{\mu}^k}{k!}\nonumber\\
 &=&\sum_{k=0}^{\infty}\frac{P_{\mu}^k}{k!}-1=\exp(P_{\mu})-1.
\end{eqnarray}
Since $P_{\mu}(t)$ is conserved, i.e $P_{\mu}(t)=P_{\mu}(0)$ for all $t\geq 0$, it follows that
\begin{equation}
 \mu \sum_{n}|\psi_n(t)|^2\le \exp(P_{\mu}(0))-1,\qquad \forall t\ge 0,
\end{equation}
and the proof is finished.\ \  $\Box$
\section{Global existence of solutions for the %IN-DNLS
Salerno lattice}
\label{secIII}
\setcounter{equation}{0}
For the current study of existence and uniqueness of a global solution of the AL and DNLS, we combine them in the so called Salerno model
% Integrable-Nonintegrable-DNLS (IN-DNLS) single system,
introduced first in \cite{Cai94}:
\begin{equation}
i\frac{d \psi_n}{dt}+(1+\mu\,|\psi_{n}|^2)(\psi_{n+1}+\psi_{n-1})+
\gamma|\psi_n|^2\psi_n=0,\,\,\,n\in {\mathbb{Z}},\label{eq:systemglobal}
\end{equation}
with $\psi_n \in {\mathbb{C}}$ and initial conditions:
\begin{equation}
 \psi_{n}(0)=\psi_{n,0},\,\,\,n \in {\mathbb{Z}}.\label{eq:icsglobal}
\end{equation}
Note that for $\gamma=0$ ($\mu=0$), the AL (DNLS) results from (\ref{eq:systemglobal}). The study of the Salerno model
%IN-DNLS
provided information about the intrinsic collapse of localised states in the presence of integrability-breaking terms, in particular, how the reflection symmetry and translational symmetry of the integrable
AL are broken by the on-site nonlinearity of the DNLS. Thereby the study of the global existence of solutions of the Salerno system
%IN-DNLS
is an essential tool in the present functional analytic set-up for the main closeness results of AL and DNLS.
%Apart from this,  this study is certainly of independent interest, due to the appearance of the non-local terms in \eqref{eq:systemglobal}.

We start by noticing that for any $\psi \in l^2$ the linear operator $A:\,l^2 \rightarrow l^2$,
\begin{equation}
 (A\psi)_{n}=\psi_{n+1}+\psi_{n-1},
\end{equation}
is continuous, since
\begin{equation}
 || A\psi||_{l^2}^2\le 4|| \psi||_{l^2}^2.\label{eq:boundlinearA}
\end{equation}
We formulate the infinite dimensional dynamical system (\ref{eq:systemglobal})-(\ref{eq:icsglobal}) as as an initial value problem in the Hilbert space $l^2$ (see \cite{NT2005}):
\begin{eqnarray}
 \dot{\psi}&=&F(\psi)\equiv i[(1+\mu\,|\psi|^2)A\psi+
\gamma|\psi|^2\psi],\,\,\,t>0,\label{eq:Hilbertsystem}\\
\psi(0)&=&\psi_0.\label{eq:Hilbertic}
\end{eqnarray}
Regarding the global existence of a unique solution to (\ref{eq:Hilbertsystem})-(\ref{eq:Hilbertic}),  we have the following
\begin{proposition}
\label{lemINDNLS}
For every $\psi_0\in l^2$,
the problem (\ref{eq:Hilbertsystem})-(\ref{eq:Hilbertic}) possesses a unique global solution $\psi(t)$ on $[0,\infty)$ belonging to
$C^1([0,\infty),l^2)$.
\end{proposition}
\noindent{\bf Proof:} First, we prove the {\it local existence} of a solution:
For this aim, the system (\ref{eq:systemglobal})-(\ref{eq:icsglobal}) is conveniently expressed as an equivalent system of integral equations
\begin{equation}
 \psi_n(t)=\psi_n(0)+i\,\int_{0}^t\left[(1+\mu\,|\psi_{n}(\tau)|^2)\Delta \psi_{n}(\tau)]+
\gamma|\psi_n(\tau)|^2\psi_n(\tau)\right] d\tau,
\end{equation}
with the notation $\Delta \psi_n=\psi_{n+1}+\psi_{n-1}$.
We consider the set
\begin{equation}
 {\cal{B}}=\left\{\phi\in C[0,\tilde{t}\,],l^2\,|\,|| \phi||_{l^2}\le \kappa \right\},
\end{equation}
which is a Banach space itself, with norm
\begin{equation}
 || \phi||_{{\cal{B}}}=\sup_{t\in [0,\tilde{t}\,]}|| \phi ||_{l^2}.
\end{equation}
Next, for $\phi \in l^2(\mathbb{Z})$, we define the nonlinear operator
\begin{equation}
 Q_n(\phi(t))=\phi_n(0)+i\,\int_{0}^t\left[(1+\mu\,|\phi_{n}(\tau)|^2)\Delta \phi_{n}+
\gamma|\phi_n(\tau)|^2\phi_n(\tau)\right]d\tau.
\end{equation}
We shall prove that the operator $Q$ establishes a contraction mapping on ${\cal{B}}$. Note first, that it satisfies the  upper bound
\begin{equation}
 || Q(\phi) ||_{{\cal{B}}}\le \kappa_0+\tilde{t}\left[(1+\mu \kappa^2)2\kappa+\gamma \kappa^3\right].
\end{equation}
We may choose $\kappa_0<\kappa/2$ and
\begin{equation}
 \tilde{t}\le \frac{\kappa}{(1+\mu \kappa^2)2\kappa+\gamma \kappa^3},
\end{equation}
so that $Q:\, {\cal{B}} \rightarrow  {\cal{B}}$. Now, for every $\phi, \psi \in  {\cal{B}}$, we have
\begin{eqnarray}
 Q_n(\phi(t))-Q_n(\psi(t))&=&i\,\int_{0}^t\left(\left[(1+\mu\,|\phi_{n}(\tau)|^2)\Delta\phi_{n}(\tau)+
\gamma|\phi_n(\tau)|^2 \phi_n(\tau)\right]\right.\nonumber\\
&-&\left.\left[(1+\mu\,|\psi_{n}(\tau)|^2)\Delta\psi_{n}(\tau)+
\gamma|\psi_n(\tau)|^2 \psi_n(\tau)\right] \right)d\tau\nonumber\\
&=&\int_{0}^t\left((\Delta \phi_n-\Delta \psi_n)+\mu (|\phi_{n}|^2\Delta \phi_n-|\psi_{n}|^2\Delta \psi_n)-\gamma (|\phi_{n}|^2 \phi_n-|\psi_{n}|^2 \psi_n)\right) d\tau,
\end{eqnarray}
and estimate the norm as
\begin{equation}
 || Q(\phi)-Q(\psi)||_{{\cal{B}}} \le \tilde{t} \left[2+(18\mu +2\gamma )\kappa^2\right]|| \phi-\psi||_{l^2}.
\end{equation}
Choosing $\tilde{t}$ such that
\begin{equation}
 \tilde{t}\le \min\left\{ \frac{\kappa}{2(1+\mu \kappa^2)\kappa+\gamma \kappa^3}, \,\frac{1}{2+(18\mu +2\gamma )\kappa^2}\right\},
\end{equation}
it is assured that $Q$ is a contraction mapping on ${\cal{B}}$. Then by Banach's fixed point theorem, there exists a {\it unique solution} of (\ref{eq:systemglobal}) provided by the unique fixed point of $Q$.
To justify that $\psi(t)$ is  $C^1$ with respect to $t$, we see from (\ref{eq:systemglobal}) that
\begin{equation}
 \sup_{t\in [0,\tilde{t}\,]} ||\dot{\psi}(t)||_{l^2} \le 2(1+\mu \kappa^2)\kappa+\gamma \kappa^3.
\end{equation}
Consequently, the solution belongs to $C^1([0,\tilde{t}\,],l^2)$.
Then, constructing a {\it maximal solution} is achieved by repeating the procedure above with initial conditions $\psi(\tilde{t}-T_{\small{f}})$ for some $0<T_{\small{f}}<\tilde{t}$.

To conclude with global existence of solutions, we remark that the Hamiltonian of \eqref{eq:systemglobal} is
\begin{eqnarray}
\label{inham}
H_{\small{S}}=\sum_n(\psi_n\overline{\psi}_{n+1}+\overline{\psi}_n\psi_{n+1})-\frac{\gamma}{2}\sum_n|\psi_n|^2-\frac{1}{\mu}\sum_n\ln(1+\mu|\psi_n|^2).
\end{eqnarray}
The \textit{deformed} Poisson-brackets are
\begin{eqnarray*}
	\{\psi_n,\overline{\psi}_m\}&=&i(1+\mu|\psi_n|^2)\delta_{nm},\\
	\{\psi_n,\psi_m\}&=&\{\overline{\psi}_n,\overline{\psi}_m\}=0,
\end{eqnarray*}	
and the equation of motion (\ref{eq:systemglobal}) is obtained as
\begin{eqnarray*}
\dot{\psi}_n=\{H_{\small{S}},\psi_n\}.
\end{eqnarray*}
The system (3.1) conserves also the quantity $P_\mu(t)$ given in \eqref{eq:Pmu} (see also \cite{Cai94}).
Then, for all initial conditions $\psi_0\in \ell^2$,
global existence in $l^2$ follows actually from the conservation of \eqref{eq:Pmu}, the help of the elementary inequality $\ln(1+\mu x)\leq \mu x$ for all $x>0$ and Lemma \ref{lemaux1}, providing that $||\psi(t)||_{l^2}
< \infty,\,\,\,\forall t\ge 0$.

%(and alternatively from \eqref{inham} which can be estimated with the help of the conservation and Lemma \ref{lemaux1}.
Similarly,  global existence for the AL \eqref{eq:AL0} is ensured by the conservation of \eqref{eq:Pmu}.
%\begin{equation}
%P_{\mu}(t)=\sum_{n}\ln(1+\mu |\psi_n(t)|^2),
%\end{equation}
For the DNLS \eqref{eq:DNLS0}, global existence in $l^2$ is established by the conservation of $P_{\gamma}(t)$ given in \eqref{eq:Pgamma}.
%\begin{equation}
% P_{\gamma}(t)=\sum_{n}|\phi_n(t)|^2.\label{eq:Pgamma}
% \end{equation}
This concludes the proof. \ \ $\Box$
%%%
\section{Proofs of Closeness of the AL and DNLS solutions}
\setcounter{equation}{0}
\label{SecIV}
\noindent{\bf Proof of Theorem \ref{Theorem:closeness}:}
Closeness will be proved in the metric ${\rm dist}_{l^2}(\varphi,\theta)=|| \varphi - \theta ||_{l^2},\,\,\forall \varphi,\theta \in l^2$. We consider the local distance of the solutions
$y_n=\phi_n-\psi_n$. On the one hand, we have that
\begin{eqnarray}
\label{dev0}
 \frac{d}{dt}|| y(t)||_{l^2}^2&=&2|| y(t)||_{l^2} \frac{d}{dt}|| y(t)||_{l^2},
\end{eqnarray}
while, on the other hand, we estimate the derivative of the $l^2$-norm as follows:
\begin{equation}
\begin{split}
\label{dev1}
  \frac{d}{dt}|| y||_{l^2}^2
  &= \sum_{n}\bigg\{i\big[(\overline{y}_{n+1}+
   \overline{y}_{n-1})y_n-({y}_{n+1}+
   {y}_{n-1})\overline{y}_n\big]\\
   &+i\mu |\psi_n|^2\big[(\overline{\psi}_{n+1}
   +\overline{\psi}_{n-1})y_n-
   ({\psi}_{n+1}
   +{\psi}_{n-1})\overline{y}_n\big]\\
   &-i\gamma |\phi_n|^2(\overline{\phi}_n y_n-\phi_n \overline{y}_n)\bigg\}\\
   &= 2\mu \sum_{n}|\psi_n|^2\bigg[(\mathrm{Im}\psi_{n+1}+\mathrm{Im}\psi_{n-1})\mathrm{Re} y_n-(\mathrm{Re}\psi_{n+1}\mathrm{Re}\psi_{n-1})\mathrm{Im} y_n\bigg]\\
   &+2\gamma \sum_{n}|\phi_n|^2\bigg[\mathrm{Im}y_n \mathrm{Re}\phi_n-\mathrm{Im}y_n \mathrm{Re}\phi_n\bigg]\\
   &\le 4\mu \sup_{n} |\psi_n|^2\sum_{n}\bigg[|\psi_{n+1}|+|\psi_{n-1}|\bigg]|y_n|+4\gamma \sup_{n} |\phi_n|^2\sum_{n}|\phi_n| |y_n|\\
   &\le 2\big(4\mu || \psi(t)||_{l^2}^3+2\gamma || \phi(t)||_{l^2}^3\big)
|| y(t)||_{l^2}.
\end{split}
\end{equation}
For the estimate \eqref{dev1}, we made use of the Cauchy-Schwarz and the continuous embeddings (\ref{eq:embeddings}).
Then, combining \eqref{dev0} and \eqref{dev1}, one has for $t>0$:
\begin{equation}
\label{ode1}
 \frac{d}{dt}|| y(t)||_{l^2}\le
 2\left(\gamma || \phi(t)||_{l^2}^3+2\mu
 || \psi(t)||_{l^2}^3\right).
\end{equation}
Note that under the hypotheses (\ref{eq:distance0})-(\ref{eq:Pmu0}), Proposition \ref{lemINDNLS} ensures that the right-hand side of \eqref{ode1} is uniformly bounded for all $t\in [0,\infty)$. Furthermore, due to the conservation of the quantities \eqref{eq:Pmu} and (\ref{eq:Pgamma}) (see also Proposition \ref{lemINDNLS}), the $l^2$ norm of $\psi$ and $\phi$ remains of the size $\epsilon$ for all times.
Integrating the inequality \eqref{ode1} in the arbitrary interval $[0, T_{\small{f}}]$, and using the assumption \eqref{eq:distance0} on the distance $||y(0)||_{l^2}=||\phi(0)-y(0)||_{l^2}$ of the initial data, we obtain that
\begin{equation*}
\begin{split}
|| y(t)||_{l^2}&\leq 2\left(\gamma C_{0,\gamma}^3+2\mu C_{0,\mu}^3\right)T_{\small{f}}\varepsilon^3+||y(0)||_{l^2}\\
&\leq  2\left(\gamma C_{0,\gamma}^3+2\mu C_{0,\mu}^3\right)T_{\small{f}}\varepsilon^3+C_0\varepsilon^3.
\end{split}
\end{equation*}
Hence, for the constant
\begin{equation}
\label{loc11}
C=2\left(\gamma C_{0,\gamma}^3+2\mu C_{0,\mu}^3\right)T_{\small{f}}+C_0,
\end{equation}
we conclude with the claimed estimate \eqref{eq:boundy}. \ \ $\Box$
%%%%%%%%%%%%%%%%%%%%%%%%%%%%%NEW
\\

The proof of Theorem \ref{Theorem:closeness} shows that the distance between the solutions of the
AL and the DNLS measured in terms of the $l^2-$metric remains small (bounded above by ${\cal{O}}(\varepsilon^3)$), compared to the $l^2-$norm of the solutions themselves.
Corollary \ref{Corollary:closeness2} follows immediately from (\ref{eq:boundy}) and continuous embedding $||y||_{l^\infty} \le ||y||_{l^2}$ and shows features for the $l^{\infty}-$norm (sup norm),
 determining the maximal distance between individual units,
 analogous to those of the $l^2-$ norm.

\section{Remarks on Extensions to higher dimensional lattices and other nonlinearities}
\label{Nlat}
\setcounter{equation}{0}
In this section, we report on extensions of Theorem \ref{Theorem:closeness} in higher dimensional lattices $\mathbb{Z}^N$, for $N\geq 2$ and for generalised nonlinearities. In the first paragraph, we comment on the closeness of the solutions of higher dimensional DNLS lattices with a generalised power nonlinearity to those of the $N$-dimensional generalisation of the AL-lattice. In the second paragraph we remark on the validity of Theorem \ref{Theorem:closeness} for the case of the DNLS with saturable nonlinearities.
\paragraph{DNLS in higher dimensional lattices.} The result of Theorem \eqref{Theorem:closeness} can be extended to higher dimensional DNLS and AL lattices of the form
\begin{eqnarray}
\label{NDNLS} i\dot{\phi}_n+(\Delta_d\phi)_n+
\gamma|\phi_n|^{2\sigma}\phi_n=0,\;\;\sigma>0,
\end{eqnarray}
and
\begin{eqnarray}
\label{NAL}
i\dot{\psi}_n+(\Delta_d\psi)_n+
\mu|\psi_n|^{2}\sum_{j=1}^N(\mathcal{T}_{j}\psi)_{n\in\mathbb{Z}^N}=0,
\end{eqnarray}
respectively.  The $N$-dimensional AL \eqref{NAL} is motivated by \cite{Kim}, where the case $N=2$ is studied as the specific limit of a 2D-generalisation of the Salerno model
%IN-DNLS
\ref{eq:systemglobal}.
The operator $(\Delta_d\psi)_n$ is the $N$-dimensional discrete Laplacian
\begin{eqnarray*}
(\Delta_d\psi)_{n\in\mathbb{Z}^N}=\sum_{m\in \mathcal{N}_n}\psi_m-2N\psi_n,
\end{eqnarray*}
where $\mathcal{N}_n$ denotes the set of $2N$ nearest neighbors of the point in $\mathbb{Z}^N$ with label $n$. With the linear operator $\mathcal{T}_{j}$ which is defined for every $\psi_n$,
$n=(n_1,n_2,\ldots,n_N)\in\mathbb{Z}^N$, as
\begin{eqnarray}
\label{defopT}
(\mathcal{T}_{j}\psi)_{n\in\mathbb{Z}^N}=\psi_{(n_1,n_2,\ldots,n_{j}+1,n_{j+1},\ldots,n_{N})}+\psi_{(n_1,n_2,\ldots,n_{j}-1,n_{j+1},\ldots,n_{N})},\;\;j=1,\ldots,N,
\end{eqnarray}
the nonlocal nonlinearity in \eqref{NAL} generalises the one of \eqref{eq:AL0}. Analytical solutions of the generalisation of the AL system \eqref{NAL} when $N=2$, have been derived in  \cite{Kim2},\cite{2DAL4},\cite{2DAL2},\cite{2DAL3}.

\paragraph{DNLS with saturable nonlinearity.}
Another important example concerns the DNLS with the \textit{saturable} nonlinearity
\begin{eqnarray}
\label{DNLS1}
\mathrm{i}\dot{\phi}_n+(\phi_{n+1}-2\phi_n+\phi_{n-1})_n+
\frac{\gamma |\phi_n|^2\phi_n}{1+\rho|\phi_n|^2}=0,\;\;\gamma,\rho>0,
\end{eqnarray}
and consequently, its other counterpart
\begin{eqnarray}
	\label{DNLS2}
	\mathrm{i}\dot{U}_n+ (U_{n+1}-2U_n+U_{n-1})-
	\frac{\Gamma U_n}{1+|U_n|^2}=0,\;\;\Gamma>0,
\end{eqnarray}
because the models \eqref{DNLS1} and \eqref{DNLS2} are not independent: solutions of the saturable model \eqref{DNLS1} can be mapped to the solutions of the model \eqref{DNLS2} by the invertible transformation
\begin{eqnarray*}
\phi_n(t)=\frac{1}{\sqrt{\rho}}\exp\left(\frac{i\gamma t}{\rho}\right)U_n(t),\;\;\Gamma=\frac{\gamma}{\rho},
\end{eqnarray*}	
see \cite{Satn1}.
For the saturable DNLS, numerous studies have verified the propagation of discrete solitons and the emergence of breathers in the 1D and 2D lattices \cite{Satn1},\cite{Satn2},\cite{Satn3},\cite{Satn4},\cite{Satn5}.
The conserved quantities of the DNLS model \eqref{DNLS2} are the power $P_{\gamma}(t)$ given in \eqref{eq:Pgamma} and the Hamiltonian
\begin{eqnarray*}
	\label{HS}
	\mathcal{H}_{\tiny{s}}=\sum_{n}|U_{n+1}-U_n|^2
	+\Gamma\sum_{n}\ln(1+|U_n|^2).
\end{eqnarray*}
The extension of Theorem \ref{Theorem:closeness} to both saturable models follows from the model \eqref{DNLS1}.
The corresponding evolution equation for the local distance $y_n=\phi_n-\psi_n$, between the solutions of  the DNLS \eqref{DNLS1} and the AL lattice \eqref{eq:AL0} is
\begin{eqnarray*}
	%\label{locNS1}
	i\dot{y}_n=-(y_{n+1}-2y_n+y_{n-1})-\left[\frac{\gamma |\phi_n|^2\phi_n}{1+\rho|\phi_n|^2}-\mu|\psi_n|^{2}(\psi_{n+1}+\psi_{n-1})\right],
\end{eqnarray*}
and the derivative of its $l^2$ norm satisfies
\begin{eqnarray}
	\label{locNS2}
	\frac{1}{2}\frac{d}{dt}|| y||_{l^2}^2=-\gamma\mathrm{Im}\sum_{n}\frac{|\phi_n|^2\phi_n}{1+\rho|\phi_n|^2}\overline{y}_n+\mu\mathrm{Im}\sum_{n}\left[|\psi_n|^{2}(\psi_{n+1}+\psi_{n-1})\right]\overline{y}_n.
\end{eqnarray}
The first term on the right-hand side of \eqref{locNS2} is estimated as
\begin{eqnarray*}
	\left|\mathrm{Im}\sum_{n}\frac{\phi_n|\phi_n|^2}{1+\rho|\phi_n|^2}\overline{y}_n\right|\leq \sum_{n}|\phi_n|^3\,|y_n|\leq ||\phi||_{l^2}^3||y||_{l^2},
\end{eqnarray*}
which can be used to derive exactly the same differential inequality \eqref{ode1}, yielding Theorem \ref{Theorem:closeness} under the same assumptions and same size of $\epsilon$ for the closeness estimate \eqref{eq:boundy}. Such an extension is also valid for higher dimensional DNLS saturable models.
\begin{figure}[tbp!]
	\begin{center}
		\begin{tabular}{cc}
			\includegraphics[width=.50\textwidth]{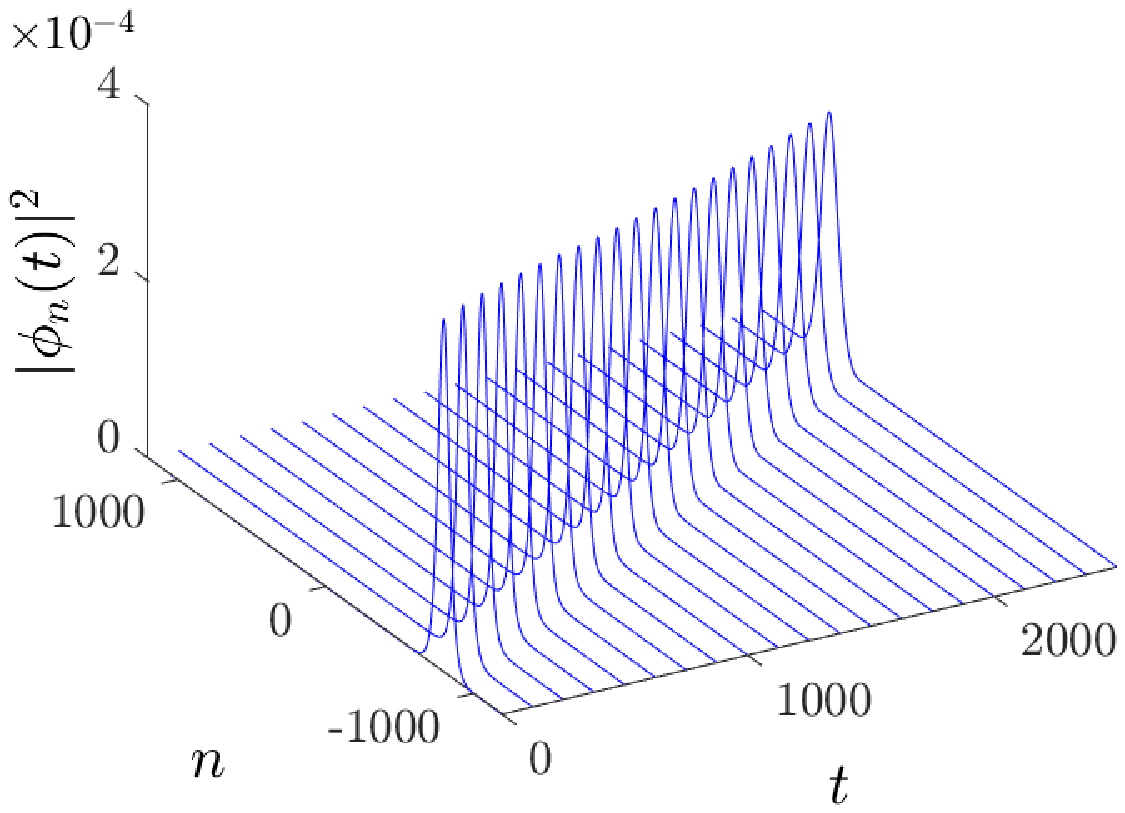}
			\includegraphics[width=.50\textwidth]{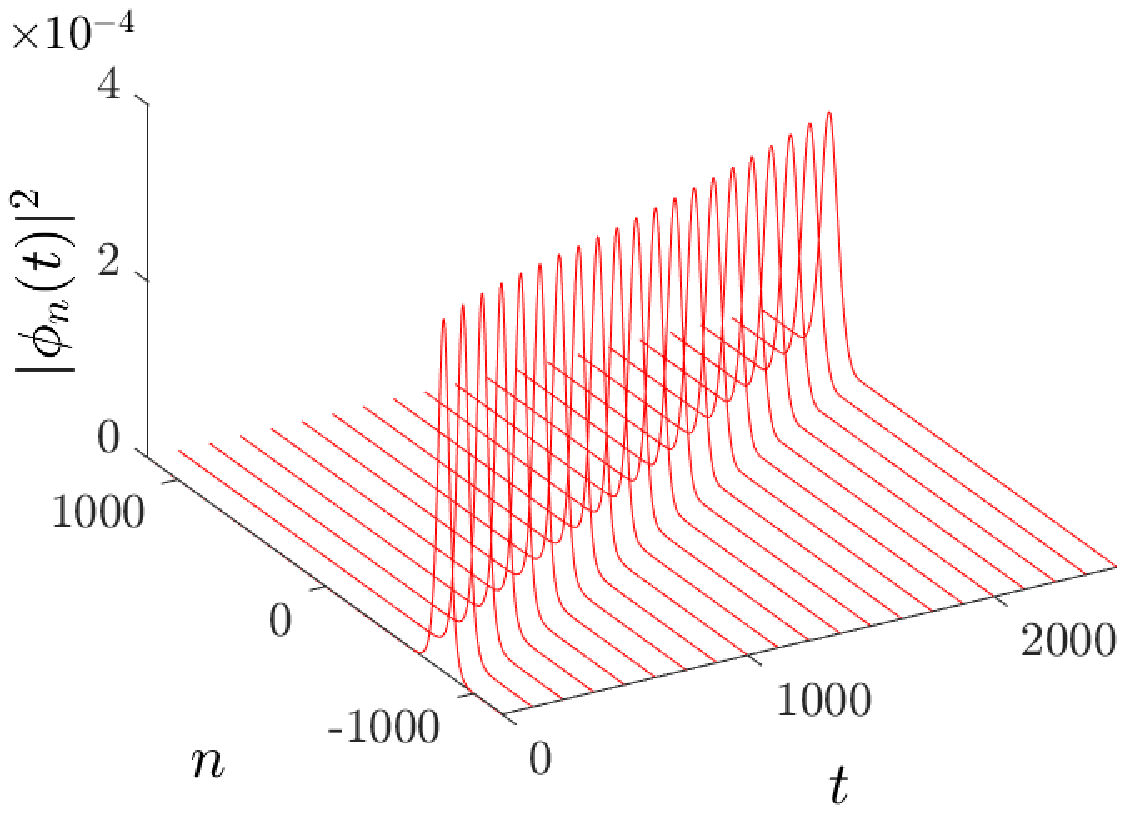}\\
		\end{tabular}
	\end{center}
	\caption{Spatiotemporal evolution of the initial condition \eqref{sin1} in the DNLS lattice with $\gamma=1$ for the cubic  nonlinearity  (left panel) and the saturable nonlinearity (right panel). Parameters of the initial condition $a=\pi/10$ and $\beta=0.02$, $\mu=1$.
	}
	\label{fig1}
\end{figure}
\begin{figure}[tbp!]
	\begin{center}
		\begin{tabular}{cc}
			\includegraphics[width=.50\textwidth]{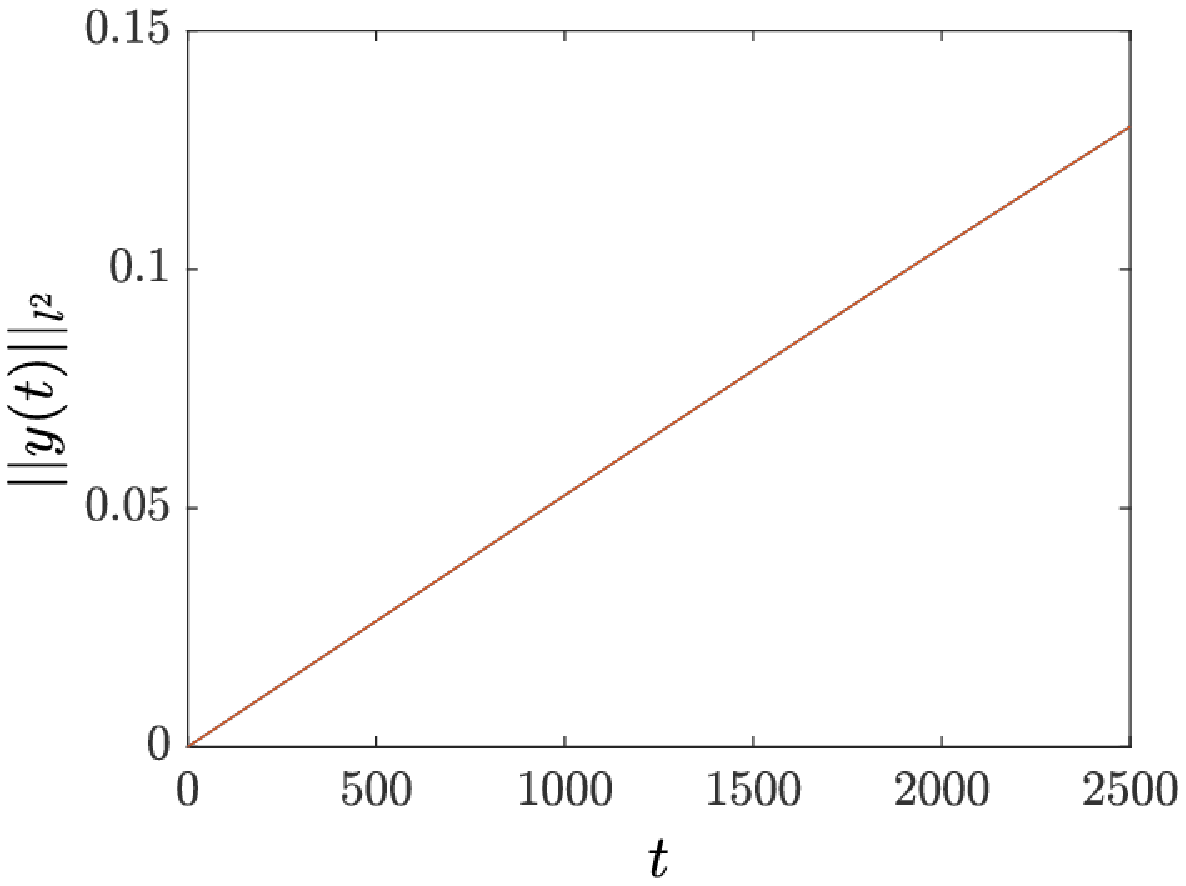}
			\includegraphics[width=.50\textwidth]{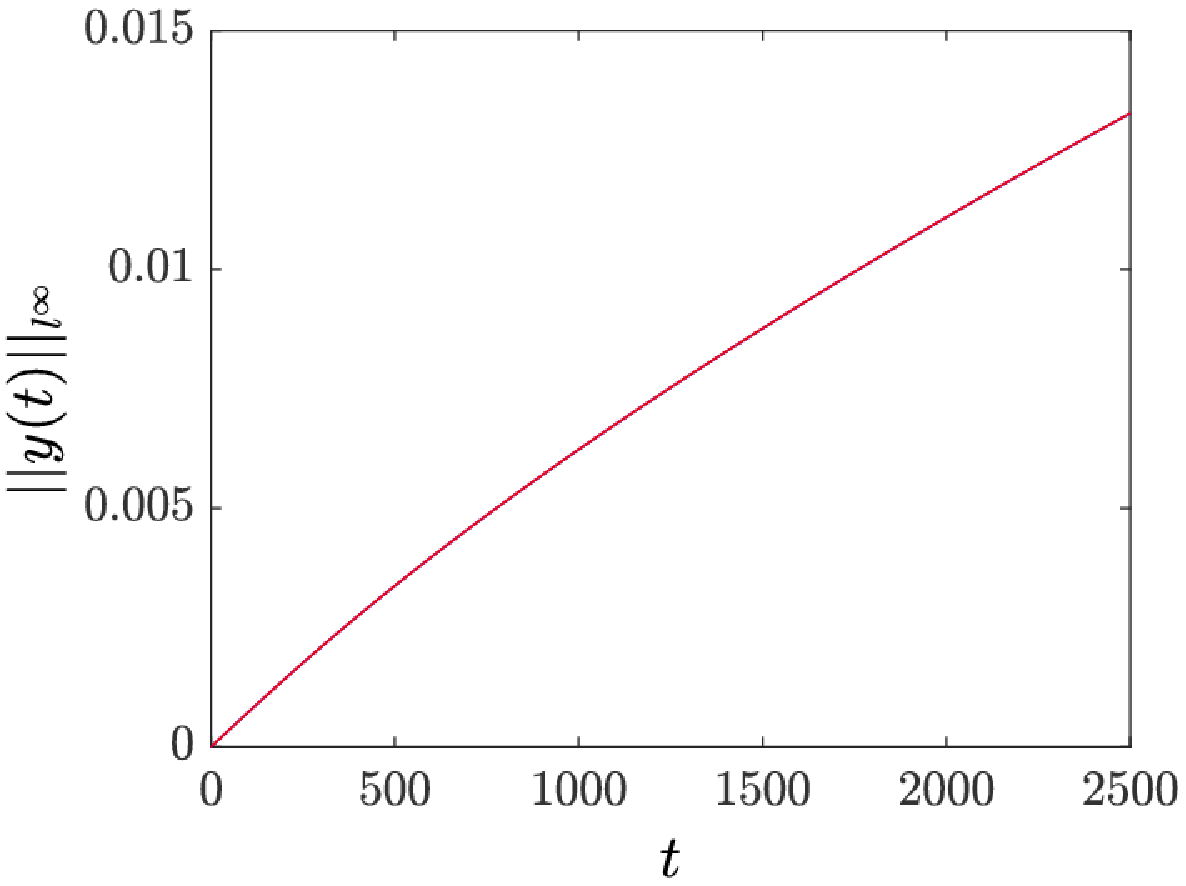}\\
			\includegraphics[width=.50\textwidth]{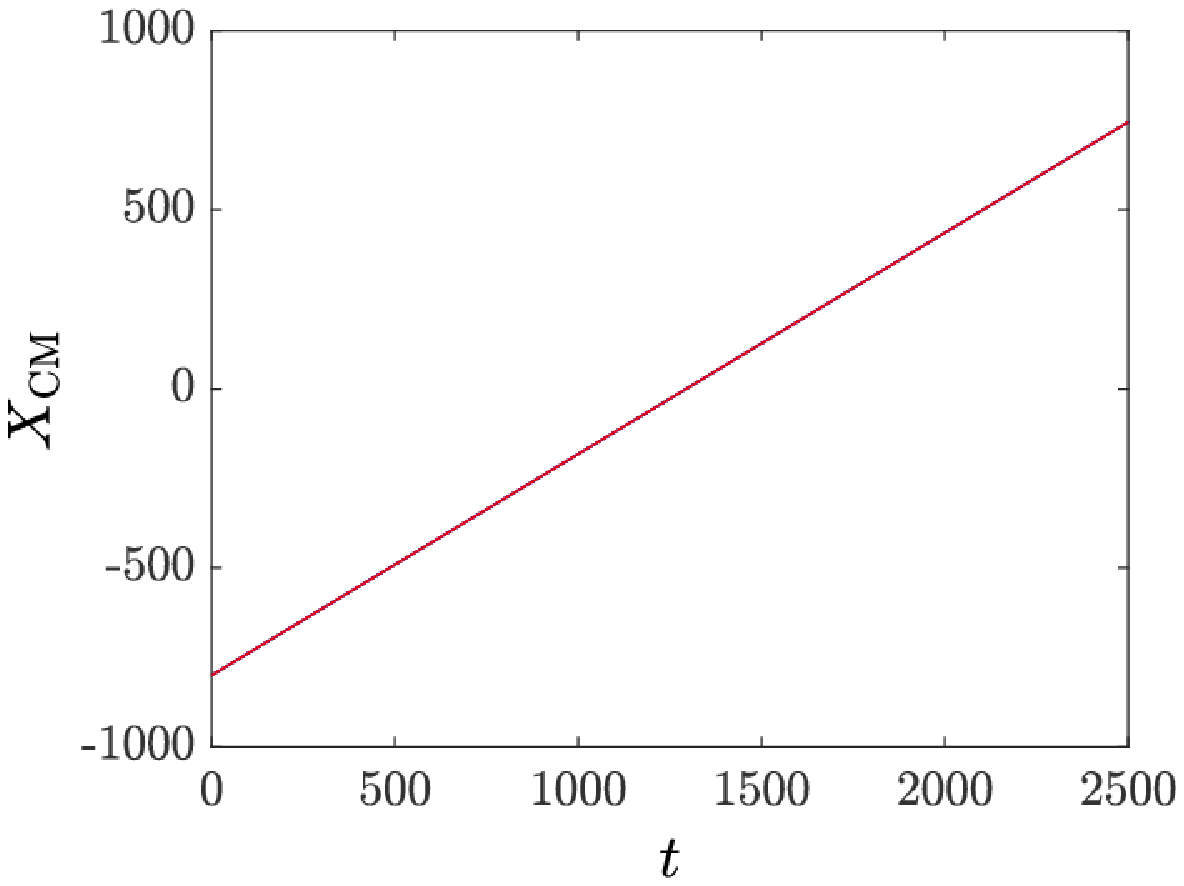}
		\end{tabular}
	\end{center}
	\caption{Top row: Time evolution of $||y(t)||_{l^2}$ and $||y(t)||_{l^{\infty}}$,  corresponding to the soliton dynamics shown in the upper panel of Figure \ref{fig1} for the DNLS with cubic and saturable nonlinearity (details in the text of section \ref{SecNum}). Bottom row: Space time evolution of the soliton center $X_{\mathrm{CM}}=(\sum_n n|\phi_n|^2)/(\sum_n |\phi_n|^2)$ for the cubic and the saturable DNLS.
	}
	\label{fig2}
\end{figure}
\begin{figure}[tbp!]
	\begin{center}
		\begin{tabular}{cc}
			\includegraphics[width=.50\textwidth]{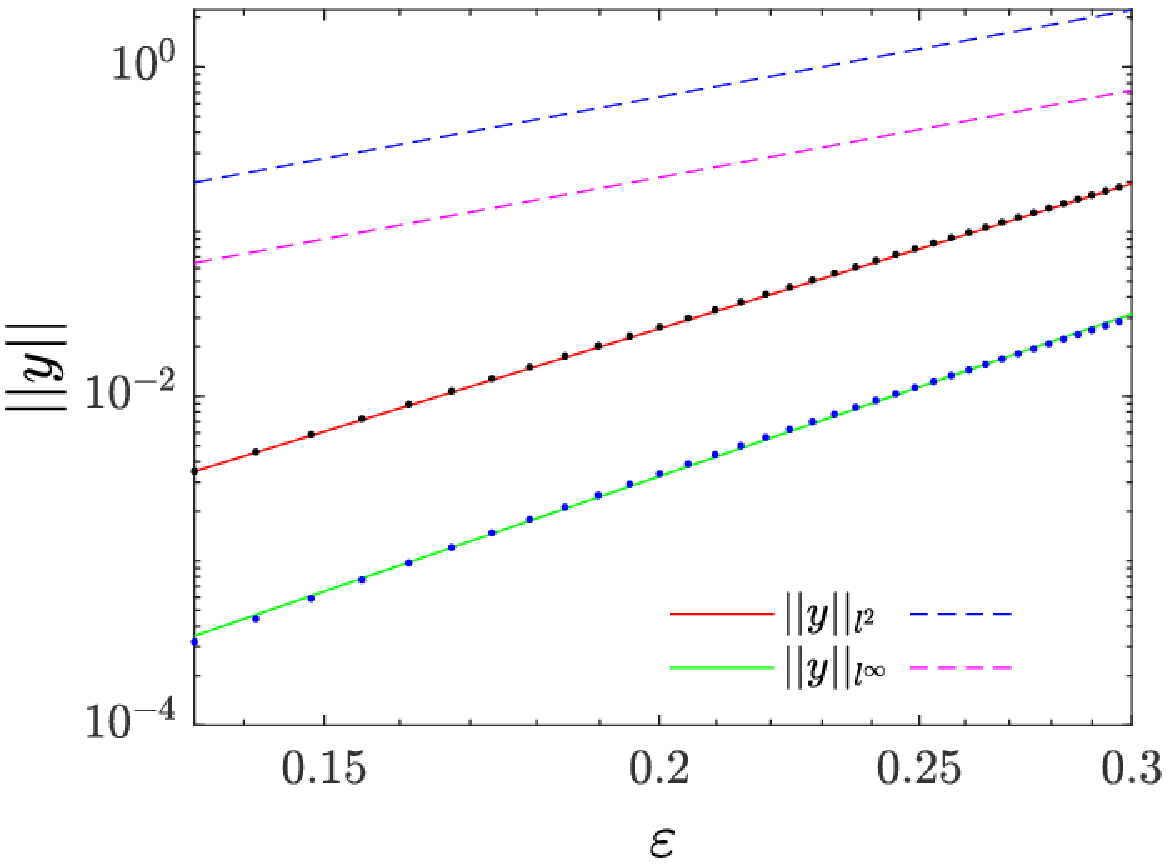}
			\includegraphics[width=.50\textwidth]{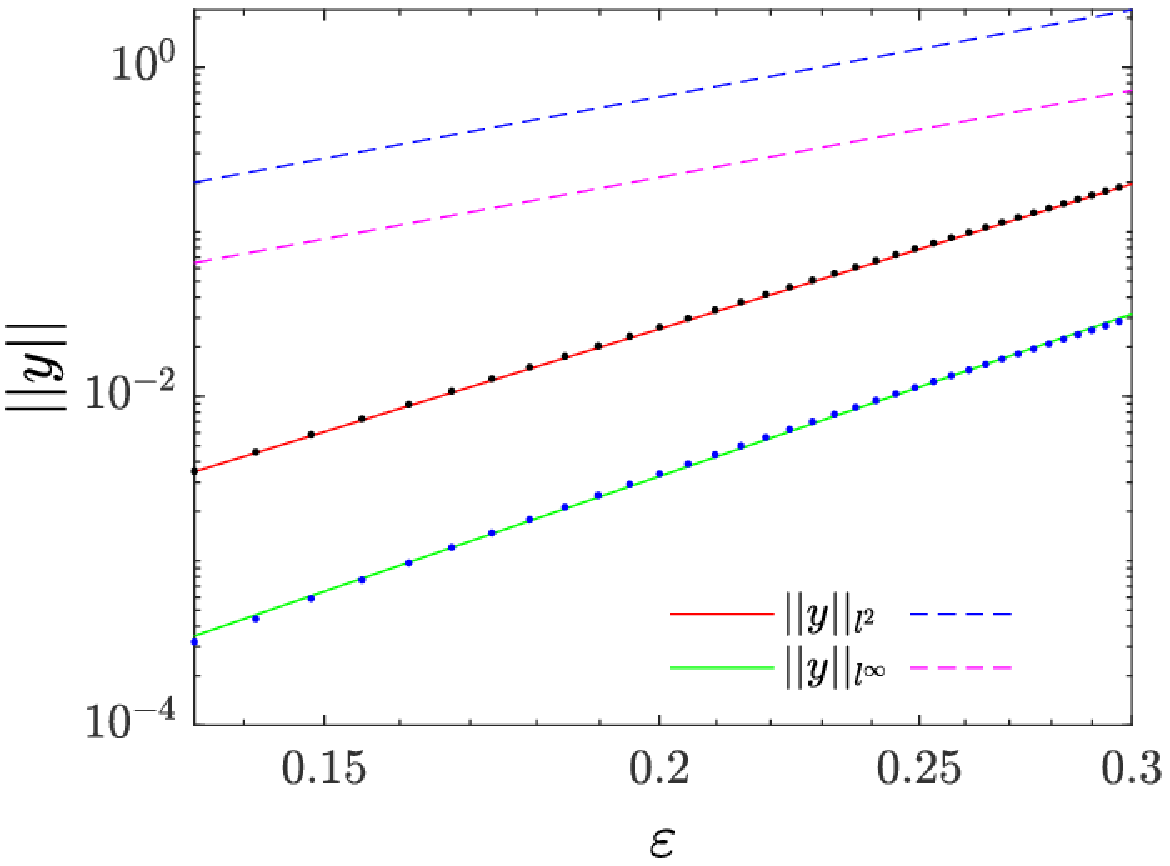}			
		\end{tabular}
	\end{center}
	\caption{ Logarithmic scaled plots of the variation of the distance functions $||y(t)||_{l^2}$  and $||y(t)||_{l^{\infty}}$ as functions of $\epsilon$, for fixed $T_f=1000$. Left panel: Cubic nonlinearity. Right panel: Saturable nonlinearity. Details are given in the text.}
	\label{fig3}
\end{figure}
\paragraph{Comments on global existence and quasi-collapse in higher dimensional lattices.} Crucial to our discussion of higher dimensional conservative  discrete nonlinear Schr\"odinger equations
is the assurance of   the global existence of their  solutions. The solutions of the DNLS \eqref{NDNLS} exist unconditionally for any $\sigma>0$ and any $N\geq 1$. This is a vital difference to its NLS pde counterpart whose solutions may blow-up in finite time when $\sigma>2/N$. The same unconditional  global existence is shared by the higher dimensional AL \eqref{NAL} or
%IN-DNLS
Salerno  models. As explained in \cite{Kim} for the 2D-lattice, if the value of the deformed power is $P_0$, then the sup-norm $||\phi||_{l^{\infty}}$ of the solution can never exceed the value $[\exp(\mu P_0)-1]$.  This is actually the argument of Lemma \ref{lemaux1}, which can be extended to the case $N\geq 1$, establishing global existence of solutions for the $N$-dimensional AL and Salerno
%IN-DNLS
equations. The work \cite{Kim} provides an analysis of the notion of \textit{quasi-collapse} which can be observed in discrete systems: Blow-up in finite time for the
the  NLS pde corresponds to the
effect of concentration of all energy in few sites in the conservative discrete NLS counterparts. The quasi-collapse supplies the mechanism for the emergence of  very narrow self-trapped states in the lattice from its initial distribution, which however, does not exhibit a finite-time-singularity (unless additional energy gain mechanisms are present \cite{Dgl}).
%%%%%%%%%%%%%%%%%%%%%%%%%%%%%%%%%%%%%%%%%%%%%%%%%%%%%%%%%555
\section{ A numerical study: Persistence of the AL-soliton in DNLS}
\setcounter{equation}{0}
\label{SecNum}
To illustrate the applicability of the analytical results we performed a numerical study examining the dynamics of the DNLS lattice for both types of nonlinearity, cubic and saturable.  As initial conditions we used the one-soliton solution of the AL \eqref{eq:one-soliton}:
\begin{eqnarray}
&&\phi_n(0)=\psi^s_n(0)\nonumber\\
\label{sin1}
&=&\frac{\sinh\beta}{\sqrt{\mu}} \mathrm{sech}(\beta n)\exp(i\alpha n),\,\,\,n\in {\mathbb{Z}},\\
&&|| \psi^s(0)||_{l^2}=|| \phi(0)||_{l^2}=\varepsilon,
\label{sin2}
\end{eqnarray}
where $\alpha \in [-\pi,\pi]$ and $\beta \in [0,\infty)$.  In order to comply with the smallness condition \eqref{sin2}, we chose the parameter values accordingly so that persistence of the corresponding  AL soliton in the DNLS can be expected.

Figure \ref{fig1} depicts the spatio-temporal evolution of the density $|\phi_n(t)|^2$ of the soliton initial condition when $\alpha=\pi/10$ and $\beta=0.02$, for the  DNLS equation with $\gamma=1$. The dynamics for the cubic (saturable) DNLS is shown in the left (right) panel. The evolution is presented for the time span $t\in [0,2500]$ ($T_{\small{f}}=2500$) and for a chain of $K=2000$ units with periodic boundary conditions. The evolution in both DNLS systems confirms the persistence of the AL soliton with amplitude of order $\mathcal{O}(\varepsilon)$ in both lattices. Notably, persistence lasts  for a significant large time interval, in particular with view to  that our analysis is a ``continuous dependence on the initial data result'' {\em where generally, the time interval of such a dependence on the initial data for a given equation might be short}. Moreover,  we studied the continuous dependence of two different systems, the cubic and the saturable DNLS, respectively.
 The dynamics of the solitons
are almost indistinguishable in both DNLS lattices. Note that for this example of initial condition, the value of the $l^2$-norm  \eqref{sin2} of the initial condition is $\varepsilon=0.2$.

Our numerical results confirm convincingly the analytical predictions presented by the Theorem \ref{Theorem:closeness} and Corollary \ref{Corollary:closeness2}, concerning the distance $$y(t)=\phi(t)-\psi(t),$$
of the solutions of the DNLS and the AL:  First, Figure \ref{fig2} depicts the time evolution of $||y(t)||_{l^2}$ (left panel) and $||y(t)||_{l^{\infty}}$ (right panel), corresponding to the dynamics of the cubic DNLS shown in the upper panel of Figure \ref{fig1}. The time evolution for the cubic DNLS is plotted as the blue curve while for the saturable as red. However, the curves are still indistinguishable, in conformity with the dynamics portrayed in Figure \ref{fig1}. The results of Figure 2, provide a first justification that both $||y(t)||_{l^2}$ and $||y(t)||_{l^{\infty}}$ remain small for significantly long time intervals.
Another interesting feature is the preservation of the soliton's speed as shown in the bottom panel of Figure \ref{fig2}. For the considered value of $\beta$, we have $\sinh(\beta)\approx\beta$, so the soliton's speed is $v_s=2\sin(\alpha)=0.6181$.

To be more precise, we examined the variation of the distances for varying small $\epsilon$. Figure 3, depicts logarithmic scaled plots of the variation of the distances  $||y||_{l^2}$ and $||y||_{l^{\infty}}$ as functions of $\epsilon$ for fixed $T_f=1000$. The left (right) panel illustrates the results of the study for the DNLS with cubic (saturable) nonlinearity.  The dashed lines in both panels, correspond to lines of the analytical estimates of Theorem \ref{Theorem:closeness} and Corollary \ref{Corollary:closeness2}, respectively,  of the form  $||y||_{l^2}$ vs  $C\epsilon^3$ and  $||y||_{l^{\infty}}$ vs $\tilde{C}\epsilon^3$. For the cubic case, we have $C=82.41, \tilde{C}=26.71$. For the saturable case, $C=82.69, \tilde{C}=26.79$. The dots on the full lines correspond to the numerically detected rates of the variations of the distance functions fitted to the lines of the form $||y||_{l^2}$ versus $C\epsilon^a$ and  $||y||_{l^{\infty}}$ versus $\tilde{C}\epsilon^b$, for the above values of constants $C$ and $\tilde{C}$: for the case of the cubic nonlinearity we found that $a=5.00$ and $b=5.60$, while for the saturable nonlinearity we found that $a=5.01$ and $b=5.60$. The numerical results illustrate that the analytical estimates are not only fulfilled, but also that the numerical variation of the distance functions is of significantly lower rate, namely of order $\sim\epsilon^5$.
%The above preliminary indicates the persistence of the AL soliton for higher amplitudes, and for  large times of numerical integration. Then, novel dynamical features emerge concerning the asymptotic state of the soliton, which will be considered in subsequent reports \cite{DJNpro}.
%%%%%%%%%%%%%%
\section{Conclusions}
\label{SecCon}
We have proved closeness of solutions of the integrable Ablowitz-Ladik equation and the non-integrable Discrete Nonlinear Schr\"odinger equation, in the sense of ``a continuous dependence of the solutions on their initial data''. For the Discrete Nonlinear Schr\"odinger equation we have considered the physically important examples of the cubic and the saturable nonlinearity. The analytical results are of relevance in regard to  the persistence of small amplitude solutions of the Ablowitz-Ladik equation in  non-integrable discrete Nonlinear Schr\"odinger equations. For an illustration of such persistence we have performed numerical simulations considering the analytical 1-soliton solution of the Ablowitz-Ladik equation. It has turned out that the  numerical findings are in excellent agreement with the analytical predictions; thus corroborating  that small amplitude solitary waves close to the analytical soliton of the Ablowitz-Ladik equation persist  in  Discrete Nonlinear Schr\"odinger equations for both types of nonlinearities, cubic as well as saturable.
Future plans shall concern studies illustrating the persistence of other localised wave forms, supplied by the analytical solutions of the Ablowitz-Ladik equation, in other (nonintegrable) Discrete Nonlinear Schr\"odinger models. An important example is that of  rational solutions. Particularly these rational solutions are non-trivial for computational studies, as the corresponding parameter values must be suitably chosen such that  "small amplitude" waveforms get formed in order to apply our analytical methods appropriate for low-amplitude solutions.  It would also be interesting to investigate the extension of the theoretical results to other models with higher order linear and nonlinear coupling operators (such as the discrete $p$-Laplacian \cite{GJ1}, the discrete biharmonic operator \cite{PNK2020}, and even other generalisations of the Ablowitz-Ladik system \cite{Kim2}, \cite{PM1}, \cite{PM2}, \cite{JC19}. Such works are in progress and relevant results will be reported
elsewhere \cite{DJNpro}.

\section*{Acknowledgment}
We would like to thank the referee for his/her constructive comments and suggestions. J.C.-M. acknowledges support from the Regional Government of Andalusia and EU (FEDER program) under the projects P18-RT-3480 and US-1380977, and MICINN, AEI and EU (FEDER program) under the projects PID2019-110430GB-C21 and PID2020-112620GB-I00.


\begin{thebibliography}{99}
\bibitem{HennigTsironis} D. Hennig and G. P. Tsironis, {\em Wave transmission in nonlinear lattices}, Phys. Rep. {\bf 307} (1999), 333--432.
%
\bibitem{Kevrekidis} P. G. Kevrekidis, K. O. Rasmussen and A. R. Bishop, {\em The discrete nonlinear Schr\"odinger equation: A survey of recent results}, Int. Journal of Modern Physics B {\bf 15} (2001), 2833--2900.
%
\bibitem{Eil} J.C. Eilbeck and M. Johansson, {The discrete nonlinear Schr\"odinger equation-20 years on} in: L. V\'azquez, R.S. MacKay, M.P. Zorzano (Eds.), {\em Localization and
	Energy Transfer in Nonlinear Systems.} World Scientific, Singapore, pp.~44--67 (2003).
%
\bibitem{DNLS} P.G. Kevredikis, {\it The Nonlinear Discrete Schr\"odinger Equation: Mathematical Analysis, Numerial Computations, and Physical Perspectives} (Springer-Verlag, Berlin, Heidelberg, 2009).
%
\bibitem{BEC1} O. Morsch and M. Oberthaler, {\em Dynamics of Bose-Einstein condensates in optical lattices}, Rev. Mod. Phys.  \textbf{78} (2006), 179--215.
%
\bibitem{AL} M.J. Ablowitz and J.F. Ladik, {\em Nonlinear differential-difference equations and Fourier analysis}, J. Math. Phys. {\bf 17} (1976), 1011--1018.
%
\bibitem{AL2} M.J. Ablowitz and J.F. Ladik, {\it A nonlinear difference scheme and inverse scattering}, Stud. Appl. Math. {\bf 65} (1976), 213--229.
%
\bibitem{Herbst} B.M. Herbst and M.J. Ablowitz, {\em Numerically induced chaos in the nonlinear Schr\"odinger equation}, Phys. Rev. Lett. {\bf 62} (1989), 2065.
%
\bibitem{Ablowitz} M.J. Ablowitz and P.A. Clarkson, {\it Solitons, Nonlinear Evolution Equations and Inverse Scattering} (Cambridge Univ. Press, New York, 1991).
%
\bibitem{Faddeev} L.D. Faddeev and L.A. Takhtajan, {\it Hamiltonian Methods in the Theory of Solitons} (Springer-Verlag, Berlin, 1987).
%
\bibitem{Ver1} V. E. Vekslerchik, {\em Functional representation of the Ablowitz–Ladik hierarchy}, J. Phys. A: Math. Gen. {\bf 31} (1998), 1087-1099.
%
\bibitem{Maruno} K. Maruno and Y. Ohta, {\it Casorati Determinant Form of Dark Soliton Solutions of the Discrete Nonlinear
	Schr\"odinger Equation}, J. Phys. Soc. Jpn {\bf 75} (2006), 054002.
%
\bibitem{akhm_AL} A.~Ankiewicz, N.~Akhmediev and J.~M.~Soto-Crespo, {\em Discrete rogue waves of the Ablowitz-Ladik and Hirota equations},
Phys. Rev. E \textbf{82} (2010), 026602.
%
\bibitem{akhm_AL2} N. Akhmediev and A. Ankiewicz, {\em Modulation instability, Fermi-Pasta-Ulam recurrence, rogue waves, nonlinear phase shift, and exact solutions of the Ablowitz-Ladik equation}, Phys. Rev. E \textbf{83} (2011), 046603.
%
\bibitem{PN1} R. Peierls, {\em The size of a dislocation}, Proc. Phys. Soc. \textbf{52}, 34--37.
%
\bibitem{PN2} R. Hobart, {\em Peierls-Barrier Minima}, J. Appl. Phys. \textbf{36} (1965), 1948--1952.
%
\bibitem{PN3} Y. Kivshar and D. Campbell, {\em Peierls-Nabarro potential barrier for highly localised nonlinear modes}, Phys. Rev. E \textbf{48} (1993), 3077--3081.
%
\bibitem{MackayAubry} R. S. Mackay and S. Aubry, {\em Proof of existence of breathers for time-reversible
or Hamiltonian networks of weakly coupled
oscillators}, Nonlinearity {\bf 7} (1994), 1623--1643.
%
\bibitem{Aubry} S. Aubry, {\em Breathers in nonlinear lattices: Existence, linear stability and quantization}, Physica D {\bf 103} (1997), 201--250.
%
\bibitem{MJ} M. Johansson and S. Aubry, {\em Existence and stability of quasiperiodic breathers in the discrete nonlinear Schr\"odinger  equation}, Nonlinearity \textbf{10} (1997), 1151--1178.
%
\bibitem{ChrisFl} J. C. Eilbeck and R. Flesch, {\em Calculation of families of solitary waves on discrete lattices}, Phys. Lett. A \textbf{149} (1990), 200--202.
%
\bibitem{AubyMarin} J. L. Marín and S. Aubry, {\em Breathers in nonlinear lattices: numerical calculation from the anticontinuous limit}, Nonlinearity \textbf{9} (1996), 1501–-1528.
%
\bibitem{Wein99} M. Weinstein, {\em Excitation thresholds for nonlinear localised modes on lattices}, Nonlinearity \textbf{12} (1999), 673--691.
%
\bibitem{Pan1} A. Pankov, {\em Gap solitons in periodic discrete nonlinear Schr\"odinger equations}, Nonlinearity \textbf{19} (2006), 27--40.
%
\bibitem{Pan2} A. Pankov, {\em Gap solitons in periodic discrete nonlinear Schrödinger equations II: a generalised Nehari manifold approach}, Discrete Contin. Dyn. Syst. \textbf{19} (2007),  419--430.
%
\bibitem{Pan3} G. Zhang and A. Pankov, {\em Standing waves of the discrete nonlinear Schr\"odinger equations with growing potentials}, Commun. Math. Anal. \textbf{5} (2008), 38--49.

\bibitem{Pan4} G. Zhang, A. Pankov, {\em Standing wave solutions of the discrete non-linear Schr\"odinger equations with unbounded potentials II}, Appl. Anal. \textbf{89} (2010), 1541--1557.
%
\bibitem{Peli0} D. Pelinovsky, {\em Translationally invariant nonlinear Schr\"odinger lattices}, Nonlinearity \textbf{19} (2006),
2695--2716.
%
\bibitem{Jesus1} T. R. O. Melvin, A.  Champneys, P. G. Kevrekidis, and J. Cuevas, {\em Travelling solitary waves in the discrete Schr\"odinger equation with saturable nonlinearity: existence, stability and dynamics}, Phys. D \textbf{237} (2008), 551-–567.
%
\bibitem{Peli1} P. G. Kevrekidis, D. E. Pelinovsky and A. Stefanov, {\em Asymptotic Stability of Small Bound States in the Discrete Nonlinear Schr\"odinger equation}, SIAM J. Math. Anal. (2009) \textbf{41}, 2010--2030.
%
\bibitem{Peli2} T. R. O. Melvin, A. R. Champneys, and D. E. Pelinovsky, {\em Discrete Traveling Solitons in the Salerno Model}, SIAM J. Applied Dynamical Systems \textbf{8} (2009), 689-–709.
%
\bibitem{FlachWillis} S. Flach and C. R. Willis, {\em Discrete Breathers}, Phys. Reports {\bf 295} (1998), 181--264.
%
\bibitem{reviewsC} S. Flach and A.V. Gorbach,
{\em Discrete Breathers -- Advances in theory and applications},
Phys. Rep. {\bf 467} (2008), 1--116.
%
\bibitem{PanosImaRev} P. G. Kevrekidis, {\em Non-linear waves in lattices: past, present, future}, IMA J. Appl. Math. \textbf{76} (2011), 389--423.
%
\bibitem{Salerno} M. Salerno, {\em Quantum deformations of the discrete nonlinear Schr\"odinger  equation}, Phys. Rev. A \textbf{46}
(1992),6856--6859.
%
\bibitem{JWU} J. Wu, {\em The Inviscid Limit of the Complex
Ginzburg- Landau Equation}, J. Differential Equations \textbf{142} (1998), 413--433.
%
\bibitem{OG} T. Ogawa and T. Yokota, {\em Uniqueness and inviscid limits of solutions for the complex Ginzburg-Landau equation in a two-dimensional domain}, Comm. Math. Phys. \textbf{245} (2004), 105--121.
%
\bibitem{Cai94} D. Cai, A. R. Bishop and N. Gr\o{}nbech-Jensen, {\em Localized States in Discrete nonlinear Schr\"odinger equation}, Phys. Rev. Lett. \textbf{72} (1994), 591--595.
%
\bibitem{Kim2} A. Khare,  K. \O{}.{} Rasmussen, M. R. Samuelsen
and A. Saxena, {\em Exact solutions of a two-dimensional
cubic–quintic discrete nonlinear Schr\"odinger equation}, Phys. Scr. \textbf{84} (2011), 065001.
%
\bibitem{2DAL4} T. Tsuchida and A. Dimakis, {On a (2+1)-dimensional generalization of the
	Ablowitz–Ladik lattice and a discrete
	Davey–Stewartson system}, J. Phys. A: Math. Theor. \textbf{44} (2011), 325206.
%
\bibitem{2DAL2} X.Y Wu, B. Tian, L. Liu and Y. Sun, {\em Bright and dark solitons for a discrete $(2+1)$-dimensional Ablowitz–Ladik equation for the nonlinear optics and Bose–Einstein condensation}, Commun. Nonlinear Sci. Numer. Simul. \textbf{50} (2017), 201--210.
%
\bibitem{2DAL3} Z.I. Djoufack, E. Tala-Tebue, J.P. Nguenang and A. Kenfack-Jiotsa, {\em Radial solitons and modulational instability in two-dimensional
Ablowitz-Ladik equation for certain applications in
nonlinear optics}, Optik \textbf{225} (2021), 165639.
%
\bibitem{NT2005} N. I. Karachalios and A. N. Yannacopoulos, {\em Global existence and compact attractors for the discrete nonlinear Schr\"odinger equation},  J. Differential Equations \textbf{217} (2005), 88--123.
%
\bibitem{Kim} P. L. Christiansen, Yu. B. Gaididei, V. K. Mezentsev ,S. L. Musher, K. \O{}. Rasmussen, J. Juul Rasmussen, I. V .
Ryzhenkova and S. K. Turitsyn, {\em Discrete Localized States and Localization Dynamics in Discrete Nonlinear Schr\"odinger Equations}, Phys. Scr. Vol. T\textbf{67} (1996), 160--166.
%
\bibitem{Satn1}
J. Cuevas and J. C. Eilbeck,
\emph{Discrete soliton collisions in a waveguide array with saturable nonlinearity},
Phys. Lett. A \textbf{358} (2006), 15--20.
%
\bibitem{Satn2}
\newblock L. Hadzievski, A. Maluckov, M. Stepic and D. Kip,
\newblock \emph{Power controlled solitons stability and steering in lattices
	with saturable nonlinearity},
\newblock Phys. Rev. Lett. \textbf{93} (2004), 033901.
%
\bibitem{Satn3}
M. Stepic, D. Kip, L. Hadzievski and A. Maluckov,
\emph{One-dimensional bright discrete solitons in media with
	saturable nonlinearity},
 Phys. Rev. E \textbf{69} (2004), 066618.
%
\bibitem{Satn4}
R. A. Vicencio and M. Johansson,
 \emph{Discrete
	soliton mobility in two dimensional waveguide arrays with
	saturable nonlinearity},
Phys. Rev. E  \textbf{73} (2006), 046602.
%
\bibitem{Satn5}
T. R. O. Melvin, A. R. Champneys, P. G. Kevrekidis and
J.  Cuevas,
\emph{Radiationless travelling waves in saturable
	nonlinear Schr\"{o}dinger lattices},
Phys. Rev. Lett. \textbf{97} (2006),
124101.
% \bibitem{Narita} K. Narita, J. Phys. Soc. Jpn. {\bf 59}, 3528 (1990).
%
\bibitem{Dgl}  G. Fotopoulos, N. I. Karachalios, V. Koukouloyannis and K. Vetas, {\em Collapse dynamics for the discrete nonlinear  Schr\"odinger equation with gain and loss}, Commun. Nonlinear Sci. Numer. Simul. \textbf{72} (2019), 213--231.
%
\bibitem{GJ1} G. James, {\em Travelling breathers and
	solitary waves in strongly
	nonlinear lattices}, Phil. Trans. R. Soc. A \textbf{376} (2018), 20170138 (25pp).
%
\bibitem{PNK2020} P. Kyriazopoulos, N. I. Karachalios and K. Vetas, {\em The Lefever-Lejeune nonlinear lattice: convergence dynamics and the structure of equilibrium states}, Phys. D \textbf{409} (2020), 132487.
%
\bibitem{PM1} M. H. Hays, C .D. Levermore, P. D. Miller, {Macroscopic lattice dynamics}, Phys. D \textbf{79} (1994), 1--15.
%
\bibitem{PM2} P. Miller, N. M. Ercolani, I. Krichever and C. D. Levremore, {\em Finite Genus Solutions to the Ablowitz-Ladik Equations}, Comm. Pure Appl. Math. \textbf{48} (1995), 1369--1440.
%
\bibitem{JC19} J. Cuevas-Maraver,  P. G.  Kevrekidis,
B. A.  Malomed and L. Guo, {\em Solitary waves in the Ablowitz–Ladik
equation with power-law nonlinearity}, J. Phys. A: Math. Theor. \textbf{52} (2019), 065202.
%
\bibitem{DJNpro} D. Hennig, N. I. Karachalios and J. Cuevas-Maraver, {\em The closeness of the integrable Ablowitz-Ladik lattice localised structures to generalised Discrete  Nonlinear  Schr\"odinger and Ablowitz-Ladik equations} (2021), in progress.


\end{thebibliography}
\end{document}